\documentclass{aipproc}
\layoutstyle{6x9}
\input psfig.sty

\newcommand\apj{ApJ}
\newcommand\apjl{ApJL}

\def\simgt{\lower.5ex\hbox{$\; \buildrel > \over \sim \;$}}
\def\simlt{\lower.5ex\hbox{$\; \buildrel < \over \sim \;$}}

\SetInternalRegister\hbadness{8000} 

\newcommand\doingARLO[2][]{%
  \ifx\mmref\undefined #1\else #2\fi
}

\begin{document}

\title 
      [Ubiquitous Relativistic Pulsar Winds]
      {Ubiquity: Relativistic Winds from Young Rotation-Powered Pulsars}


\author{E. V. Gotthelf}{
  address={Columbia Astrophysics Laboratory, 550 West 120th St, New York, NY 10027, USA},
  email={eric@astro.columbia.edu},
  thanks={This work made possible by NASA LTSA grant NAG~5-7935}
}

\copyrightyear  {2001}

\begin{abstract}
Recent X-ray observations of young rotation-powered pulsars are
providing an unprecedented detailed view of pulsar wind nebulae. For
the first time, coherent emission features involving wisps, co-aligned
toroidal structures, and axial jets are fully resolved in X-rays on
arc-second scales.  These structures, which are remarkably coherent
and symmetric, are similar to features seen in the optical Crab
nebula.  In this report, we present the latest high resolution Chandra
X-ray images of six young rotation-powered pulsars in supernova
remnants. These data suggest an X-ray morphology perhaps common to all
young rotation-powered pulsars
and serve as a guide for developing
the next generation of theoretical models for pulsar wind nebulae.

\end{abstract}

\date{\today}

\maketitle

\section{Introduction}


Young Crab-like pulsars are thought to lose their rotational energy
predominantly in the form of a highly relativistic particle
wind. Evidence for this wind is manifest as a bright, centrally
condensed synchrotron emission nebula, which Weiler \& Panagia (1978)
referred to as a ``plerion''.
Models for these pulsar wind nebulae (PWNe) are based on the Crab
Nebula, the first known and brightest example.  The freely expanding
pulsar wind is initially invisible as it travels though the
surrounding self-evacuated region. Eventually, the wind encounters the
ambient medium where it is reverse-shocked, resulting in the
thermalization and re-acceleration of particles. Synchrotron radiation
from these particles is most easily observed in the form of a bright
radio nebula which acts as a calorimeter of the pulsar's current
energy loss. Models were developed by Pacini and Salvati (1973),
Reynolds \& Chevalier (1984) and others and subsequently refined
(e.g. Kennel \& Coroniti 1984; Aron 1998).  A
review is found in Chevalier (1998).

High resolution Chandra X-ray images of the Crab pulsar show the
limitations of the current models. These observations display coherent
concentric toroidal structures and jet-like features apparently
aligned along the spin axis, and in the direction of the pulsar's
velocity vector. The remarkable structures and alignments clearly
delineate the complex magneto-hydrodynamics associated with this
pulsar which have yet to be fully explained.  Herein we present
several more examples of Crab-like pulsars in supernova remnants
(SNRs) observed with Chandra, all of which show evidence of similarly
orientated toroidal and possible jet-like features. Collectively,
these observations suggest, for the first time, the fundamental
relationship of these structures to the central engines in young
rotation-powered pulsars. We briefly present each pulsar and discuss
the implications of their individual and common morphology.

\section{New Chandra Observations}

The Chandra Observatory (Weisskopf, O'Dell, and van Speybroeck 1996)
has targeted most of the known Crab-like pulsars using one or both of
its two imaging focal plane detectors, the High Resolution Camera
(HRC) and the CCD-camera (ACIS). Both cameras provide arc-sec imaging
over a $\sim 0.5^{\circ} \times 0.5^{\circ}$ field-of-view. The HRC
allows pulse-phase imaging to isolate the pulsars from their nebulae
while ACIS provides moderate resolution spectroscopy.  Table I
presents a summary of the observational characteristics of the pulsars
presented herein. The Chandra images for each of these SNRs are
displayed in Fig. 1a-3b.


\medskip
{\noindent \bf PSR~J0534$+$2200, The Crab [Fig. 1a]} -- The remarkable
Chandra observation of the Crab nebula has been reported in Weisskopf
et. al 2000.  As previously mentioned, images from this data set shows
toroidal and jet-like X-ray structures aligned along the spin
axis. These features were first hinted at by earlier X-ray observation
(Aschenbach \& Brinkmann 1975) but the quality of the new Chandra
images now fully reveals the central engine powering the nebula, 
clearly delineating its geometry with respect to the optical nebula.
The observed morphology displayed in Fig. 1a forms the basis for
a comparison with other PWN pulsars. The Crab Nebula, however, is unique
amongst young supernova remnants in that it has no discernible SNR
shell; the reason for this is not yet satisfactorily explained.

\begin{figure} 
\centerline{ {\hfil\hfil
\psfig{figure=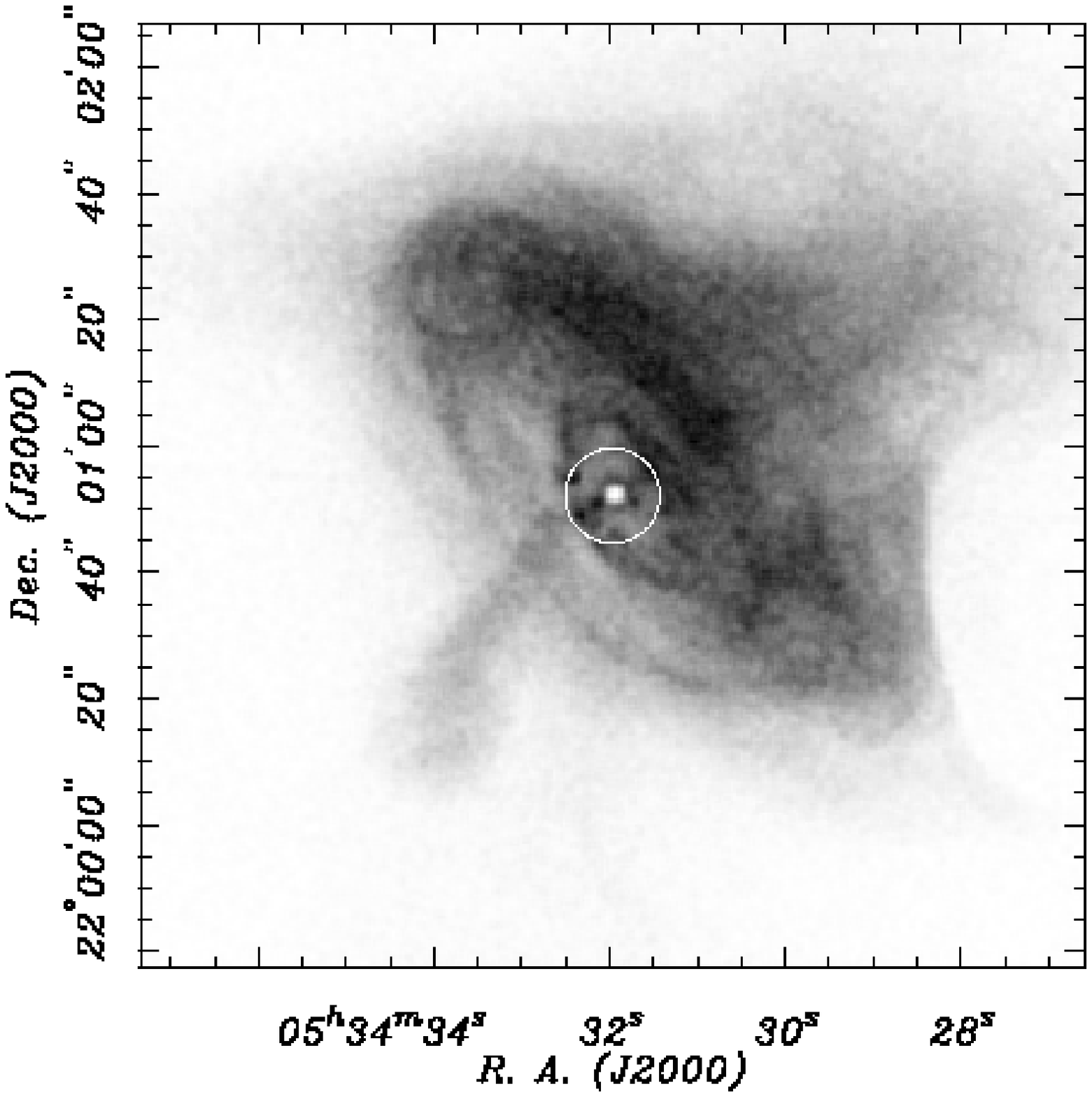,height=2.8in,angle=270, clip=}
\quad
\psfig{figure=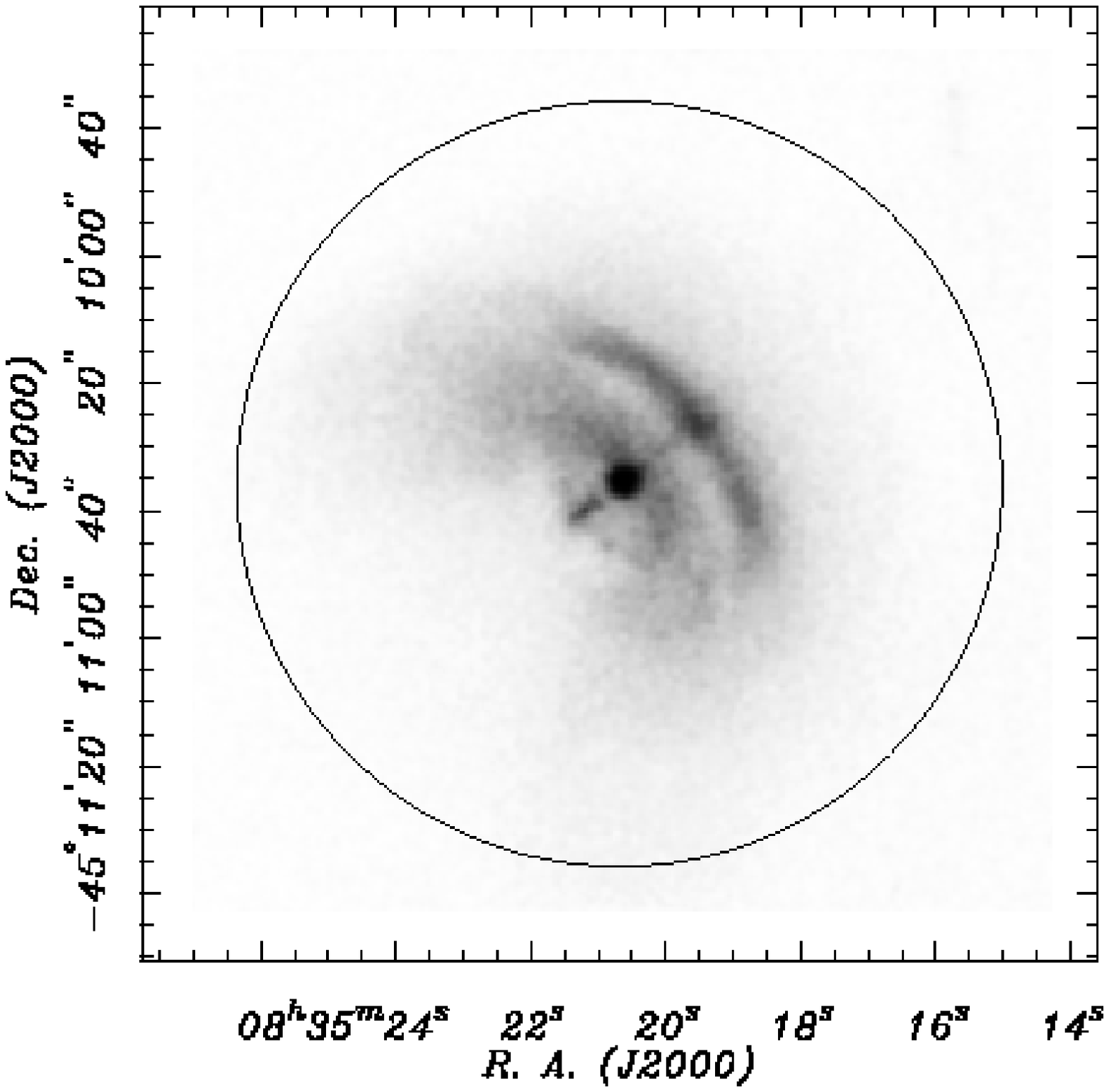,height=2.8in,angle=270, clip=}
\hfil\hfil} }
\caption{A scaled comparison of the relativistic wind nebulae
surrounding two young pulsars observed by the Chandra Observatory,
(left) the 1,000 yrs Crab pulsar and (right) the 12 kyrs Vela
pulsar. These images are displayed with the same plate scale, but the
Vela nebula is a factor of 16 times smaller than the Crab assuming
distances of 2 kpc (Crab) and 250 pc (Vela); the circles represent the
same physical size at the distance of the pulsar. Although Vela is an
order of magnitude older and smaller, the two objects are found to be
similar in shape and overall brightness distribution. From Helfand,
Gotthelf, \& Halpern (2001); see also Weisskopf (2000)}
\end{figure}

\medskip
{\noindent \bf PSR~J0835$-$4510 in Vela~XYZ [Fig. 1b]} -- The 89~ms
Vela pulsar was observed twice with the HRC, $\sim 3.5$ and $\sim 35$
days following an extreme glitch in its rotation frequency (Helfand,
Gotthelf, and Halpern 2001). Most surprisingly was the discovery of a
coherent toroidal structure and axial jet remarkably similar to that
observed in the Chandra observation of the Crab Nebula. Furthermore,
as for the Crab, the axis of symmetry is also along the directions of
the pulsar proper motion. Clearly resolved are two concentric arcs
which may form tori. The physical size of these structures is 16 times
smaller relative to the Crab (see Figure 1), perhaps commensurate with
its lesser spin-down energy.  Comparison of the two Vela observations
shows that the brightness of the outer arc increases significantly while
the flux of the pulsar remained relatively steady. If this increase is
associated with the glitch, the inferred propagation velocity is
$\simgt 0.7c$, similar to that seen in the brightening of the optical
wisps of the Crab Nebula.

\begin{figure} 
\centerline{ {\hfil\hfil
\psfig{figure=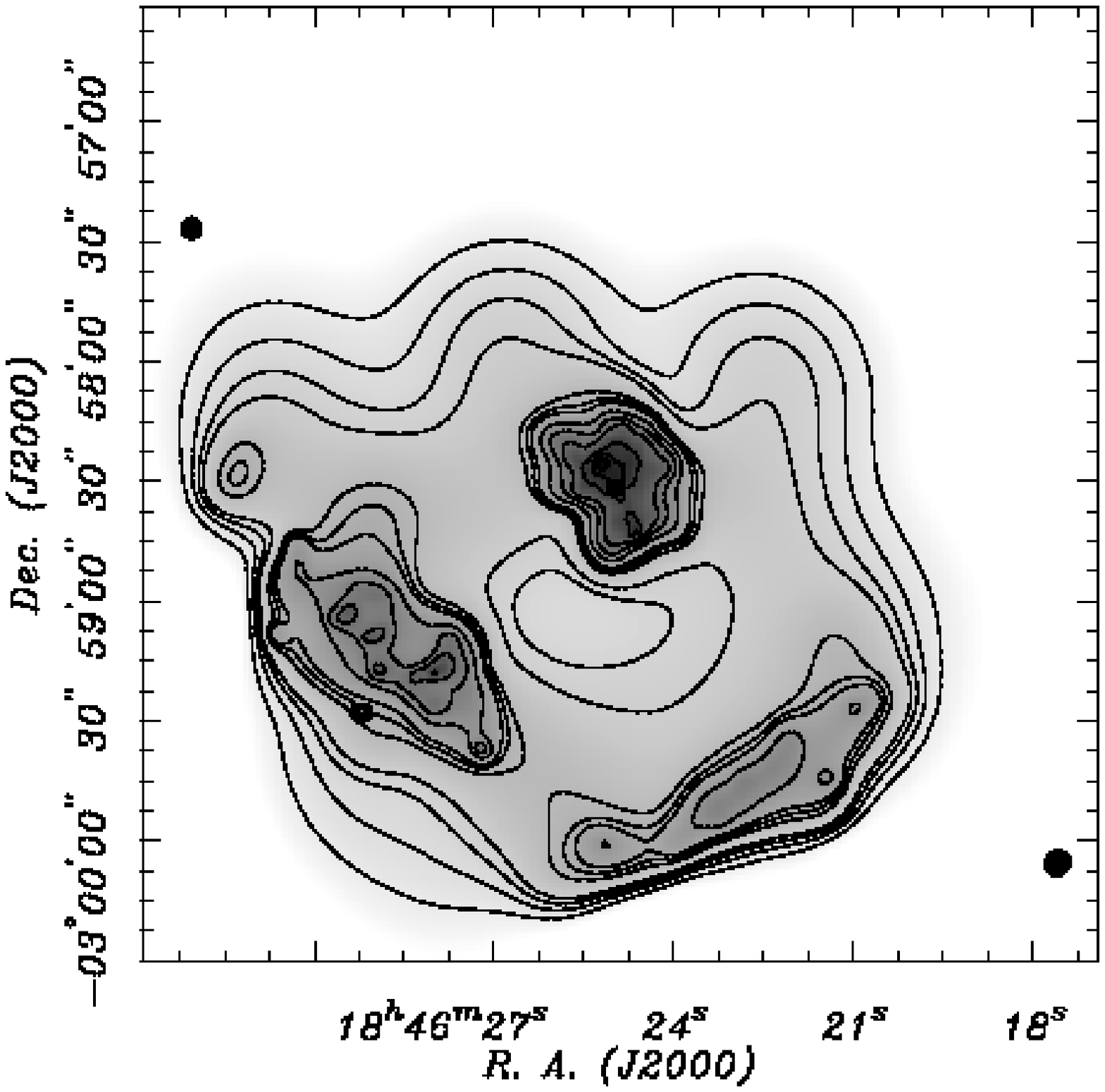,height=2.8in,angle=270, clip=}
\quad
\psfig{figure=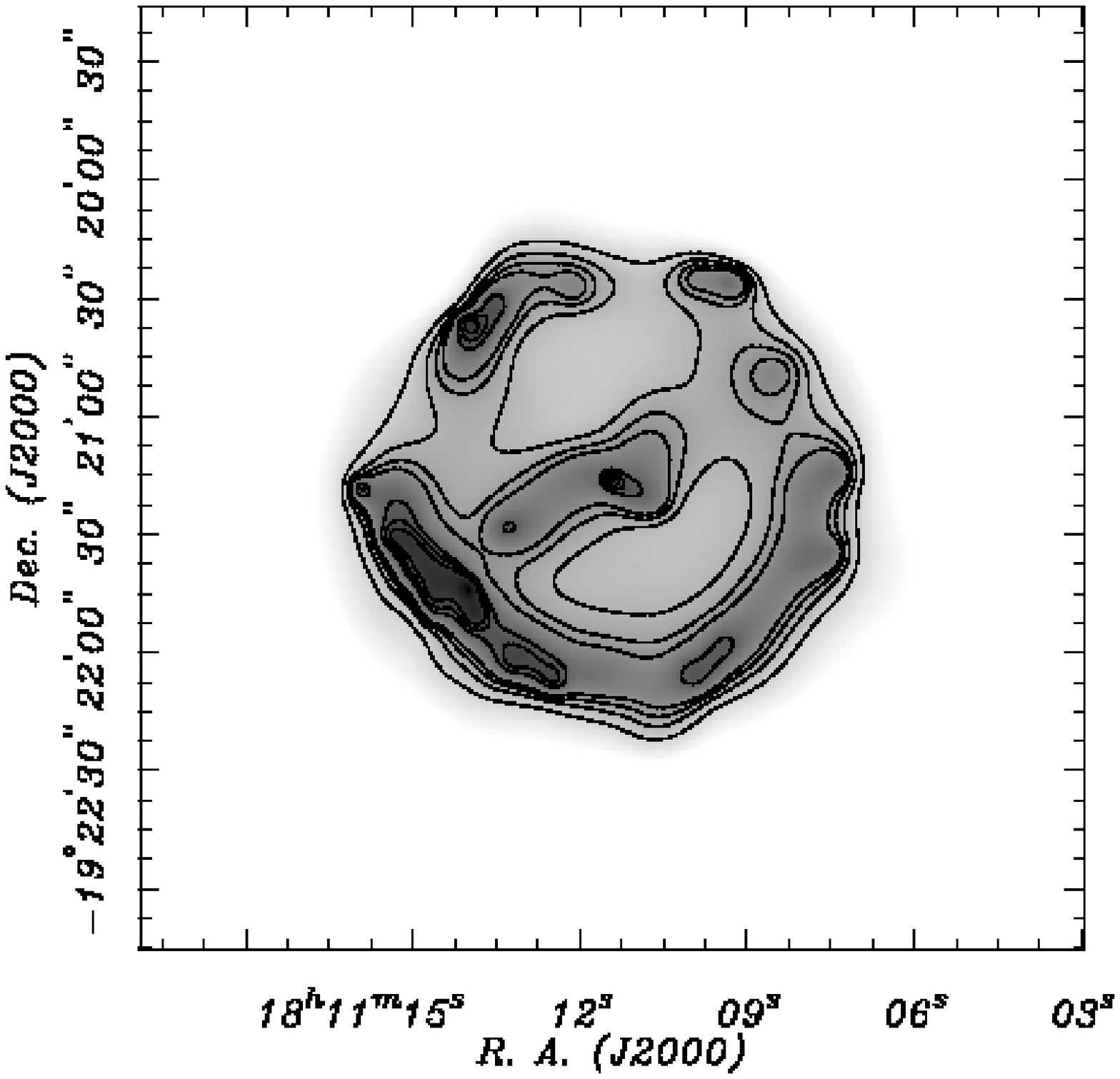,height=2.8in,angle=270, clip=}
\hfil\hfil} }
\caption{ Two young supernova remnants observed by the Chandra
Observatory containing recently discovered pulsars. (Left) -- Kes 75
with its 324 ms pulsar PSR~J1846$-$0258, which has a spin-down age of
only 700 yrs. (Right) -- G11.2-0.3 contains the 69 ms pulsar
PSR~J1811$-$1926, possibly the stellar remnant of the historic
supernova of A.D. 386. The resolved pulsars in the center of both
images are by far the brightest object in the image which is saturated
to highlight the diffuse emission. These images are
adaptively smoothed and their respective contour levels and plate
scales set the same for comparison. From Helfand \& Gotthelf (2001)
and Kaspi et al. (2001)}
\end{figure}

\medskip
{\noindent \bf PSR~J1846$-$0258 in Kes~75 [Fig. 2a]} -- Kes~75 is a
young, distant ($\sim 19$ kpc) Galactic shell-type remnant with a
central core whose observed properties have long suggested a PWN
similar to the Crab Nebula (Becker, Helfand \& Szymkowiak 1983).
Located within the core of Kes~75 resides the recently discovered
PSR~J1846$-$0258 (Gotthelf et al. 2000) -- a pulsar with exceptional
timing properties: its period, spin-down rate, and spin-down
conversion efficiency, are each an order-of-magnitude greater than
that of the Crab, most likely as a result of its extreme magnetic
field (Helfand \& Gotthelf 2001). Although the PWN is found to be
noticeable elongated, the statistics of the current observations is
insufficient to resolve any detailed Crab-like structure. The
association of a shell-type remnant in Kes~75 with a centrally
located, coeval pulsar provides strong evidence that neutron stars are
born in supernovae explosions. PSR~J1846$-$0258 has the youngest
characteristic age ($P/2\dot P \sim 700$ yrs) and is likely being spun
down rapidly by torques from a large magnetic dipole, just above
$B_{QED}$ (see Table 1). The role of the magnetic field is important for
understanding the transport and dissipation of particle and magnetic
field energy in the relativistic pulsar wind.

\medskip
{\noindent \bf PSR~J1811$-$1926 in G11.2$-$0.3 [Fig. 2b]} -- The
apparently young SNR G11.2$-$0.3 is a shell-type Galactic remnant
containing the 65 ms pulsar PSR~J1811$-$1926 (Torii et al 1998). Based
on it location and thermal remnant age estimate, G11.2-0.3 has been
proposed as the remnant of supernova SN~386, one of only a few
historical supernovae known. The spin-down age of PSR~J1811$-$1926 is,
however, 24,000 yrs, a severe discrepancy which suggests the two
sources could be unrelated. This is most puzzling, as the pulsar is
located near the geometric center of the nearly complete X-ray and
radio shell. This suggests the intriguing possibility that the pulsar
was born spinning near its current rate or suffered an episode of
rapid spin-down.  If associated, the symmetry and completeness of the
thermal shell strongly constrains the projected magnitude of any
initial velocity imparted to the pulsar during its formation.

\begin{figure} 
\centerline{ {\hfil\hfil
\psfig{figure=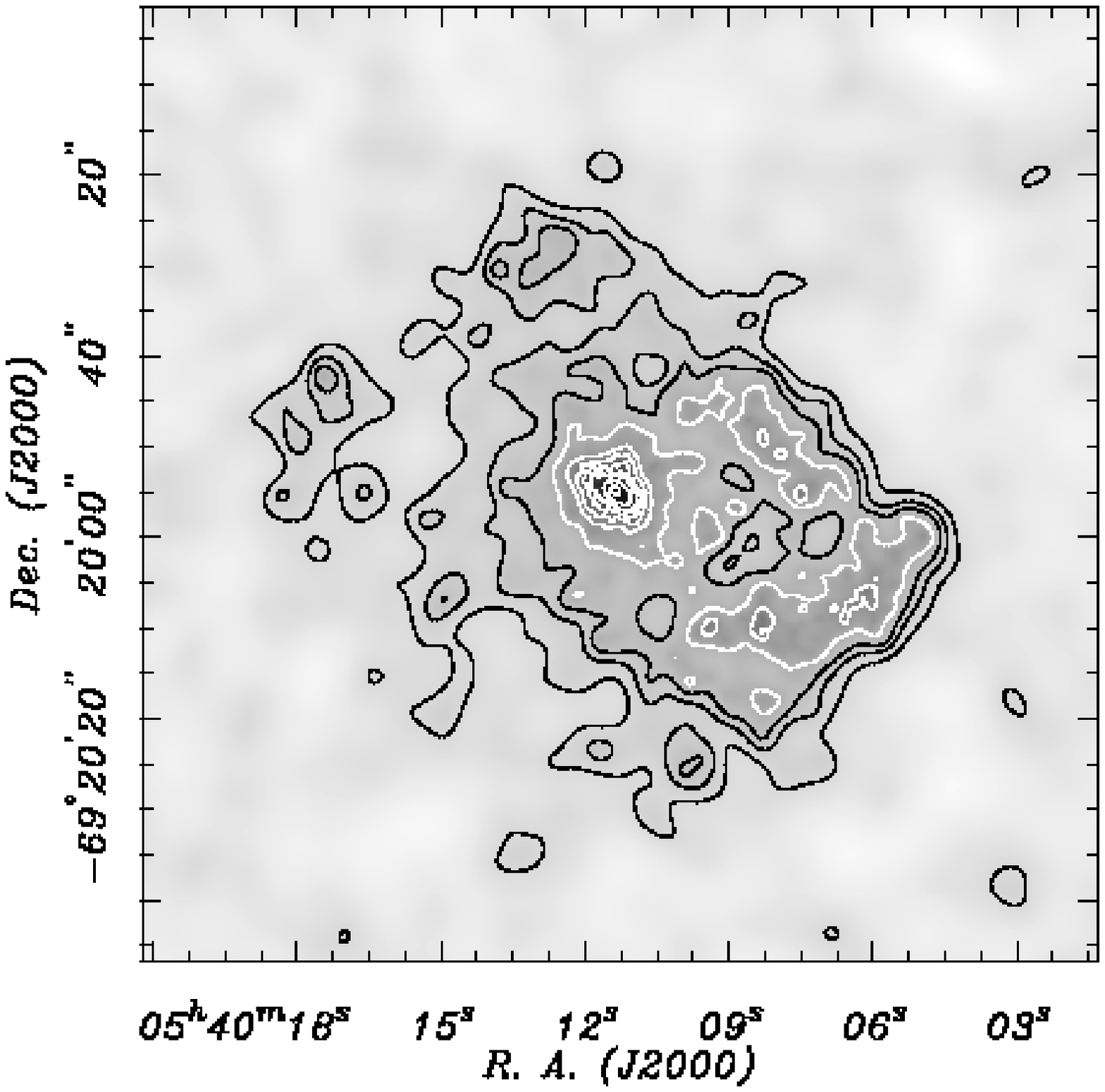,height=2.8in,angle=270, clip=}
\quad
\psfig{figure=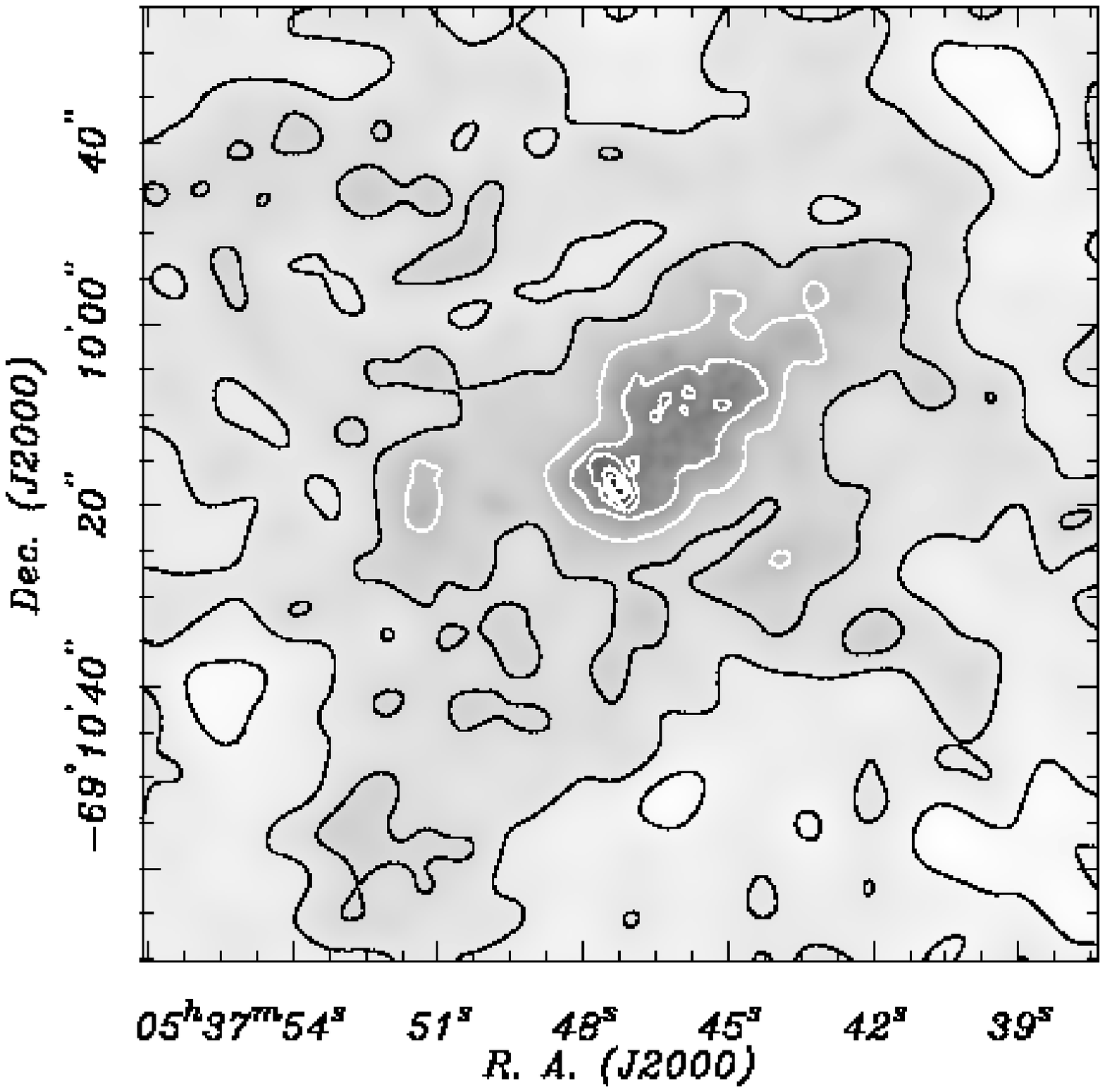,height=2.8in,angle=270, clip=}
\hfil\hfil} }
\caption{A global view around two young Crab-like pulsars located
close to each other in the Large Magellenic Clouds. (Left) -- the 50
ms pulsar PSR~J0540$-$6919 in SNR~0540-69.3 has an incomplete SNR
shell containing an elongated core. Phase resolved blow-up of the core
(see Gotthelf \& Wang 2000) shows weak evidence of a perpendicular
jet. (Right) -- the recently discovered 16 ms pulsar PSR~J0537$-$6910
in the N157B nebula, the most rapidly spinning young pulsar known, has
faint SNR emission, a bright PWN with an enormous tail of diffuse
emission.  These images are adaptively smoothed and their respective
contour levels and plate scales set the same for comparison.  From
Gotthelf \& Wang (2000) and Wang \& Gotthelf (2001).}
\end{figure}

\medskip
{\noindent \bf PSR~J0540$-$6919 in SNR~0540-69.3 [Fig. 3a]} -- The Chandra
calibration observation of the X-ray-bright 50 ms pulsar
PSR~J0540$-$6919 in the Large Magellanic Cloud (LMC) has conclusively
demonstrated its Crab-like nature (Gotthelf \& Wang 2000; Kaaret et
al. 2000). Although much older than the Crab (based on its spin-down
age), PSR~J0540$-$6919 is contained within a central PWN with toroidal
structure similar in size to that of the Crab. The PWN is just
resolved at the distance of the LMC, but analyses of pulse
phase-dependent images reveals the characteristic elliptical emission
of a Crab-like torus, plus weak evidence for a jet emanating from the
pulsar, perpendicular to the major axis of the torus as seen for the
Crab and Vela systems.

\medskip
{\noindent \bf PSR~J0537$-$6910 in N157B [Fig. 3b]} -- This remnant is
also located in the LMC, only 17 arcmins from SNR~0540-69.3, and contains the
recently discovered 16 ms pulsar PSR~J0537-6910, the most energetic
and rapidly rotating young pulsar known. In addition to the pulsar
itself and its surrounding compact PWN, the Chandra X-ray observations
resolve a third distinct features, a region of large-scale diffuse
emission trailing from the pulsar (Wang, Gotthelf, Chu, \& Dickel
2001). This X-ray feature, the largest among all known Crab-like SNRs,
is a comet-shaped bubble coexisting with enhanced radio emission and
is oriented nearly perpendicular to the major axis of the PWN. It is
likely powered by a toroidal pulsar wind of relativistic particles
which is partially confined by the ram-pressure from the supersonic
motion of the pulsar.  Ram-pressure confinement also allows a natural
explanation for the observed X-ray luminosity of the compact nebula
and for the unusually small X-ray to spin-down luminosity ratio of
$\sim 0.2\%$, compared to the Crab pulsar.

\begin{table}
\begin{tabular}{lrrrrrrr}
\hline
  \tablehead{1}{c}{b}{Remnant\tablenote{References for Chandra images -- G11.2$-$0.3: Kaspi et al. (2001); Vela~XYZ: Helfand, Gotthelf, \& Halpern (2001); Kes~75: Helfand \& Gotthelf (2001); SNR~0540-69.3: Wang \& Gotthelf (2000), Kaaret et al. (2000); Crab Nebula: Weisskopf et. al (2000); N157B Nebula: Wang, Gotthelf, Chu, \& Dickel (2001).}}
  & \tablehead{1}{c}{b}{Pulsar}
  & \tablehead{1}{c}{b}{Distance \\ (kpc)}   
  & \tablehead{1}{c}{b}{P \\(ms)}   
  & \tablehead{1}{c}{b}{$\bf \dot P$ \\ $\times 10^{-14}$ \\ (s/s)}   
  & \tablehead{1}{c}{b}{Age \\ (kyrs)}   
  & \tablehead{1}{c}{b}{$ \bf \dot E$\tablenote{Ranked by spin-down energy.} \\ $\times 10^{35}$ \\ (ergs/s)}   
  & \tablehead{1}{c}{b}{B$_p$/B$_{QED}$\tablenote{The inferred magnetic field is normalized to the quantum critical field defined 
as B$_{QED} = m^2_e c^3 / e \hbar = 4.4 \times 10^{13}\ {\rm G.}$} }   
\\
\hline
G11.2$-$0.3  & PSR~J1811$-$1926 & 5 & 69  & $4.4$   & 23.0  &   63& 0.04\\
Vela~XYZ     & PSR~J0835$-$4510 & 0.25   & 89  & $12.0$  & 12.0  &   67& 0.08\\
Kes~75       & PSR~J1846$-$0258 & 19& 324 & $710.0$ & 0.7 &   82& 1.10\\
SNR~0540-69.3 & PSR~J0540$-$6919 & LMC   & 50  & $48.0$  & 17.0  & 1516& 0.11\\
Crab~Nebula  & PSR~J0534$+$2200 & 2 & 33  & $40.0$  & 1.3 & 4394& 0.08\\
N157B~Nebula & PSR~J0537$-$6910 & LMC   & 16  & $5.1$      & 5.0 & 4916& 0.02\\
\hline
\end{tabular}
\caption{Young Crab-like Pulsars in Supernova Remnants: 
Observational Characteristics}
\label{tab:a}
\end{table}

\section{Discussion}

This preliminary look at the new Chandra images of rotation-powered
pulsars hint at what might be learned from these observations. Common
to these pulsars is their location in the centers of their respective
SNR. Apparently these pulsars have not travelled far from their
origin. Collectively, these images provide important constraints on
the average birth-kick velocity imparted to young pulsars.

The conjecture that the Crab Nebula is the result of the historic
supernova SN~1054 provides a direct link between neutron stars
formation and supernovae explosions. The case may be better made,
however, by pulsars residing in identifiable thermal SNR shells,
apparently absent for the Crab. 
The 1000-yr young Crab may be in a pre-shell evolutionary stage;
however, a counter example is provided by Kes~75, whose pulsar has a
similar spin-down age coeval with its SNR. Furthermore, N157B and
G11.2$-$0.3 show only weak shell emission, perhaps like the Crab, but
their characteristic ages are much older. As often suggested, the
spin-down age is likely unreliable for young pulsars. This notion is
furthered advanced by the association of G11.2$-$0.3 with SN~386,
which is inconsistent with the spin-down age of PSR~J1811$-$1926. The
standard assumptions used to associate the spin-down age with the ``true''
pulsar age may well be violated for young pulsars.

Perhaps ultimately more revealing is the common alignment of the
symmetry axis for several of the PWN tori and jets along the direction
of the pulsar's velocity vector, when known.  Quite inscrutable is the
ostensible chance sky alignment of the principle axis for the observed
PWNs with position angle of $\sim 45$ degs. What is clear, however, is
that the observed complex toroidal and jet structures are likely
ubiquitous to young rotation-powered pulsars. Furthermore,
imaging-spectroscopy of these feature are consistent with a highly
energetic particle wind emitting non-thermal synchrotron radiation.
Of course much theoretical work is needed to model these remarkable
structures.

\begin{theacknowledgments}
This work is made possible by NASA LTSA grant NAG~5-7935.
\end{theacknowledgments}

\hyphenation{Post-Script Sprin-ger}

\end{document}